\documentclass[aps,pra,floatfix,tightenlines,amsmath,amssymb,twocolumn,showpacs]{revtex4}

\usepackage{graphicx}
\usepackage{amssymb}
\usepackage{color}
\usepackage{psfrag}
\usepackage{ifsym}
\usepackage{hyperref}
\usepackage{braket}

\definecolor{pink}{rgb}{1,0.078,0.57}

\newcommand{\ee}[1] {\mathrm{e}^{#1}}

\newcommand{\dg}{^{\dagger}}

\newcommand{\hc}{\mathrm{h.c.}}

\begin{document}
\title{Self-calibrating tomography for multi-dimensional systems}
\author{Nicol\'as Quesada$^1$, Agata M. Bra\'nczyk$^2$ and Daniel F. V. James$^1$}
\email{nquesada@physics.utoronto.ca}
\address{$^1$ McLennan Physical Laboratories, University of Toronto, 60 St. George Street, Toronto, ON, M5S1A7, Canada.\\
$^2$Department of Chemistry and Center for Quantum Information and Quantum Control, 80 Saint George Street, University of Toronto, Toronto, Ontario, M5S 3H6 Canada }

\date{\today}
\begin{abstract}
Quantum state tomography relies on the ability to perform well-characterized unitary operations, which are not always available. This places restrictions on which physical systems can be characterized using such techniques. We develop a formalism that circumvents the requirement for well-characterized unitary operations by treating unknown parameters in the state and unitary process on the same footing; thereby simultaneously performing state and partial process tomography. Our formalism is generalized to $d$-level systems, and we provide a specific example for a V-type three-level atomic system whose transition dipole moments are not known.  We show that it is always possible to retrieve not only the unknown state but also the process parameters, except for a set of zero measure in the state-parameter space. 
\end{abstract}


\maketitle

\section{Introduction}
The characterization of quantum states and processes is a key step in quantum information processing and quantum computing \cite{Nielsen2000}. For quantum states, such characterization can be achieved by performing a variety of measurements on many identically-prepared copies of the state, then using these results to reconstruct the density matrix that represents the state. This technique is known as  \emph{quantum state tomography} (QST)  \cite{Vogel1989,Leonhardt1995,James01}. 

The diverse set of prescribed measurements is typically achieved by applying unitary transformations to the quantum state before measurement in a convenient basis.  This requires a well-characterized unitary operation, which is not always available. \emph{Self-calibrating tomography} (SCT) aims to circumvent this requirement by treating all unknown parameters---both in the state and the process---on the same footing \cite{aggie11}. SCT allows the reconstruction of a quantum state despite incomplete knowledge of the unitary operations used to change the measurement basis, while simultaneously solving for the unknown parameters in the unitary. 

The duality between unknown parameters in the state and process was first highlighted by Brif and Mann \cite{Brif2000} in the context of spectroscopy and interferometry, and the concept of performing tomography with unknown parameters in both the state, as well as the process, was introduced by Mogilevtsev   \cite{Mogilevtsev2010} in the context of detector tomography (see also subsequent work by Mogilevtsev \emph{et al.} \cite{Mogilevtsev2012}). Related ideas have also been investigated in the context of gate set tomography by Merkel \emph{et al.} \cite{Merkel2012}. 

The use of partially characterized unitary operations for quantum state tomography was first considered by Bra\'nczyk \emph{et al.}  \cite{aggie11}. There we examined a specific example: tomography of single and entangled qubits using unitary operations with one unknown parameter. We successfully reconstructed the state of polarization-encoded photonic qubits using a wave plate with unknown retardance.  In this work we extend and formalize the ideas introduced in \cite{aggie11} and provide a full theoretical treatment in Section \ref{qubits}. Using this formalism, we then show how to generalize SCT to the case of multi-dimensional systems in Section \ref{sec:gen}. The characterization of multi-dimensional systems is relevant to fundamental questions regarding non-locality in higher dimensions \cite{Kaszlikowski2002,Collins2002} as well as improved security using quantum key distribution \cite{Bruss2002,Cerf2002}. Recent interest in quantum effects in biological systems \cite{Engel2007,Collini2010,Panitchayangkoon2010}---where complex molecular systems undergo poorly-characterized evolution---further motivates the need for techniques such as SCT for higher-dimensional systems. 

To demonstrate the adaptation to higher-dimensional systems, we consider a specific physical example in Section \ref{sec:vlev}: a non-degenerate V-type three-level atomic system with unknown transition dipole moments, coupled to the radiation field. This system is a basic model for the strongly-coupled dimer found in cryptophyte antenna proteins \cite{Collini2010}. We show how SCT can be used to recover the quantum state of the three-levels system as well as determine the unknown transition dipole moments. 

\section{Two-level system}\label{qubits}

In this section, we develop a full theoretical treatment for self-calibrating tomography (SCT) in the context of qubits. We explore several scenarios including the one originally discussed in \cite{aggie11} and we prove that such schemes will be able to reconstruct  any possible state of a qubit system except for a set of zero measure. In this exception, a different scheme can be used.

A requirement of quantum state tomography is the ability to perform measurements in a variety of different bases. In many physical systems, there is a ``preferred'' basis in which a measurement can be made, e.g. the horizontal/vertical polarization basis in polarization-encoded single photons; or the ground/excited state basis for atomic systems. Other bases are typically accessed by performing well-characterized unitary operations on the state before measurement, effectively changing the measurement basis. 

Self-calibrating tomography considers the case where this unitary operation is not completely characterized, yet controllable to some extent. It is ``self-calibrating'' in the sense that certain parameters defining the measurement basis do not need to be calibrated in advance.

To illustrate the concept, we begin by defining a general qubit density operator and a general generator of rotations that is related to the Hamiltonian $\hat{H}(t)$ that governs the unitary evolution of the state. We then show how the measurement statistics depend on the parameters of the initial state and the unitary. 

The single-qubit state $\hat\rho$ and generator of rotations  under which the state evolves $\hat G=\frac{1}{\hbar}\int dt\hat{H}(t)$ can be written in the eigenbasis of the Pauli operator $\hat \sigma_z$, i.e. $\{\ket{0},\ket{1}\}$, as follows:
\begin{subequations}\label{eq:rhoH}
\begin{align}\label{par1}
\hat\rho={}&\left(
\begin{array}{cc}
 \rho _{00} & e^{-i \gamma } \rho _{01} \\
 e^{i \gamma } \rho _{01} & \rho _{11}
\end{array}
\right)\,;\\\label{par1b}
\hat G ={}&\frac{1}{2}\left(
\begin{array}{cc}
 h_z & e^{-i \phi } h_{c} \\
 e^{i \phi } h_{c} & -h_z
\end{array}
\right).
\end{align}
\end{subequations}

This parametrization yields simpler analytical expressions for the measurement statistics than the usual normalized Stokes' parameters. In addition, it highlights the relationship between the angles $\phi$ and $\gamma$, which will be discussed below. The components of the Bloch vectors associated with $\hat \rho$ and $\hat G$ are given by:
\begin{subequations}\label{bloch}
\begin{align}
\vec v_{\hat \rho}&={}\left\{2 \rho _{01} \cos (\gamma ),2 \rho _{01} \sin (\gamma ),\rho _{00}-\rho_{11}\right\}\,;\\\label{eq:vH}
\vec v_{\hat G}&={}\left\{h_{c} \cos (\phi ),h_{c} \sin (\phi ),h_z\right\}\,.
\end{align}
\end{subequations}

For $\hat \rho$ to provide a valid description of a quantum state, it must have positive eigenvalues and unit trace. Given the parametrization in Equation (\ref{par1}), these constraints are equivalent to 
\begin{align}
|\vec v_{\hat \rho}|= \sqrt{4 \rho _{01}^2+\left(\rho _{00}-\rho _{11}\right){}^2}\leq \text{tr}(\hat \rho) = \rho _{00}+\rho _{11} =1\,.
\end{align}

The action of the Hamiltonian on the state will be given by
\begin{equation}\label{action}
\hat \rho \longrightarrow \hat \rho_{\nu} =  \mathcal{\hat U_{\nu}} \hat \rho  \mathcal{\hat U_{\nu}}^\dagger\,,
\end{equation}
with
\begin{equation}\label{u}
\mathcal{\hat U_{\nu}}(h_z,h_{c},\phi)=\ee{-i \hat G (h_z,h_{c},\phi)}\,,
\end{equation}
where we assume that the Hamiltonian commutes with itself at all times. We use the label $\nu$ to refer to a specific set of parameters $\{h_z,h_{c},\phi\}$ within the generator $G$. 

\begin{figure*}
\begin{center}
\includegraphics[width=\textwidth]{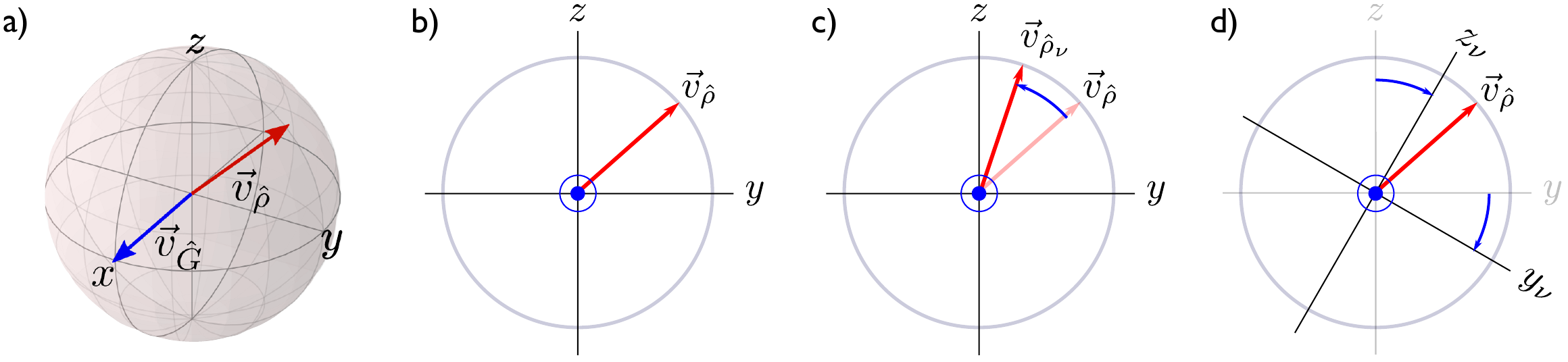}\end{center}
\caption{The Bloch vectors $\vec v_{\hat \rho}$ and $\vec v_{\hat G}$, given in Equation (\ref{bloch}), shown on: (a) the Bloch sphere; and (b) a cross-section of the Bloch sphere. The unitary operation $\mathcal{\hat U_{\nu}}$, shown in Equation (\ref{u}), can be thought of as either: (c) rotating $\vec v_{\hat \rho}$ about $\vec v_{\hat G}$; or (d) rotating the reference frame about $\vec v_{\hat G}$ in the opposite direction.}
\label{rotations} 
\end{figure*}

By expanding the exponential it is easy to see that the Bloch vector of the state $\hat \rho$ is rotated by an angle 
\begin{align}\label{eq:omega}
\Omega=|\vec v_{\hat G}|=\sqrt{h_{c}^2+h_z^2}
\end{align}
around the axis specified by the Bloch vector of $\hat G$. This is then an active transformation of the vector. 
 Equivalently the action of the Hamiltonian can be defined in terms of the basis elements $\{\ket{0}, \ket{1} \}$:
\begin{subequations}\label{action2}
\begin{align}
\ket{0}\longrightarrow \ket{0_{\nu}}&={}\mathcal{\hat U_{\nu}}^\dagger \ket{0}\,;\\
\ket{1}\longrightarrow \ket{1_{\nu}}&={}\mathcal{\hat U_{\nu}}^\dagger \ket{1}. 
\end{align}
\end{subequations}

In this interpretation, the Bloch vector of the state remains fixed but the Bloch sphere itself rotates by an angle ${-}\Omega$ around the axis $\vec v_{\hat G}$. This is the passive transformation of the reference frame, which is equivalent to the transformation of the basis vectors and is illustrated in FIG. (\ref{rotations}). Either way, transformation of $\hat\rho$ according to $\mathcal{\hat U}_{\nu}$, followed by measurement in the $\{\ket{0},\ket{1}\}$ basis constitutes measurements given by the projectors
\begin{align}
\hat\Pi_{\nu}^{0}={}&\ket{0_{\nu}}\bra{0_{\nu}};~~~\hat\Pi_{\nu}^{1}={}\ket{1_{\nu}}\bra{1_{\nu}} \,.
\end{align}

We draw attention to the fact that the superscript $j$ in $\hat\Pi_{\nu}^{j}$ is a label denoting the original projector and should not be mistaken for an exponent. The statistics associated with measuring $\hat\Pi_{\nu}^{0}$ are given by
\begin{align}\label{bitstats0}
n^0_{\nu}={}&\text{tr}\left(\hat \rho' \hat\Pi_{\nu}^{0}\right)=\sum_{i<j} f_{ij} \rho_{ij}\,,
\end{align}
where $\hat \rho'=\mathcal{N}\hat \rho $ and $\mathcal{N}$ is a constant that depends on the duration of data collection, detector efficiency, loss etc. In what follows we reabsorb this constant into $\hat \rho'$ and thus recast the condition $\text{tr}(\hat \rho)=1$ into $\text{tr}(\hat \rho')=\mathcal{N}$. Once the unnomarlized density matrix $\hat \rho'$ is determined it suffices to divide by its trace to recover $\hat\rho$. The coefficients in Equation (\ref{bitstats0}) are given by
\begin{subequations}\label{functionsf}
\begin{align}
f_{00}&={}c^2+s^2 \tilde h_z^2\,;\\
f_{11}&={}s^2 \tilde h_c^2\,;\\
f_{01}&={}2  \left(c \sin (\beta )+s \cos (\beta ) \tilde h_z\right)\sqrt{f_{11}}\,,
\end{align}
\end{subequations}
where $\tilde h_i ={} h_i/\Omega$, $c={}\cos( \Omega/2)$, $s={}\sin(\Omega /2)$, and $\beta=\phi-\gamma$. We note that there is a non-linear relationship between the measurement statistics $n^0_{\nu}$ and the parameters $h_z$,  $h_c$ and $\beta$. Similarly, the statistics associated with measuring $\hat\Pi_{\nu}^{1}$ are given by
\begin{align}\label{bitstats1}
n^1_{\nu}={}&\text{tr}\left(\hat \rho' \hat\Pi_{\nu}^{1}\right)=\sum_{i<j} g_{ij} \rho_{ij}\,,
\end{align}
where
\begin{subequations}\label{functionsg}
\begin{align}
g_{00}&={}s^2 \tilde h_c^2\,;\\
g_{11}&={}c^2+s^2 \tilde h_z^2\,;\\
g_{01}&={}-2  \left(c \sin (\beta )+s \cos (\beta ) \tilde h_z\right)\sqrt{g_{00}}\,.
\end{align}
\end{subequations}

We now examine the forms of $n_{\nu}^{0}$ and $n_{\nu}^{1}$ to determine how much information can be extracted about the state and the process. It can be seen that just one type of measurement, either  $\hat\Pi_{\nu}^{0}$ or $\hat\Pi_{\nu}^{1}$, is sufficient to access the elements $\rho_{00}$, $\rho_{01}$ and $\rho_{11}$  of the density matrix in Equation (\ref{par1}).   Inspection of the coefficients $f_{01}$ and $g_{01}$ shows that the system of equations can also be solved for the elements $h_c$ and $h_z$ in the unitary in Equation (\ref{u}). Further inspection reveals that the system of equations can only be solved for $\beta=\phi-\gamma$:  only if the phase $\phi$ is known, can the complete state be reconstructed; similarly, only if the phase $\gamma$ is known, can the complete unitary be reconstructed. 

We now define a set of measurement statistics $\Lambda\subseteq\{{n}_{\nu_1}^{0},{n}_{\nu_2}^{0}\dots,{n}_{\nu_1}^{1},{n}_{\nu_2}^{1}\dots\}$ and a set of unknown parameters $\Gamma\subseteq\{\rho_{00},\rho_{01},\rho_{11}, h_{z},h_{c}, \beta\}$. Determining whether the set of equations in (\ref{bitstats0}) and/or (\ref{bitstats1}), parameterized by $\Lambda$, will be invertible for $\Gamma$ is a standard problem of analysis \cite{Jacobi96}. This occurs when the Jacobian, i.e. the determinant of the Jacobian matrix, is non-zero, 
\begin{align}\label{eq:jac}
J=\text{det}\left(\frac{\partial \Lambda}{\partial \Gamma} \right) \neq 0\,.
\end{align}

It is important to mention that when the inversion is performed in practice, errors will be present in the measurement statistics. For ``small'' values of the Jacobian, the inversion might not always be completely meaningful. The study of errors is important and will need to be done on a case-by-case basis, but lies outside the scope of this paper.

An appropriate set of different realizations of one of the projectors  $\hat\Pi_{\nu}^{j}$ that satisfies the invertibility condition in Equation (\ref{eq:jac}) can be used to construct a set of equations invertible for the unknown parameters, thereby realizing quantum state tomography of the two-level system and/or quantum process tomography of the unitary.

In the remainder of this section, we discuss several scenarios, beginning with a completely characterized unitary before extending the formalism to include unknown parameters in the generator $\hat G$. 

\subsection{Completely characterized unitary}

We first consider the familiar scenario where one has complete knowledge and control over the unitary operation. Given the ability to measure in one particular basis, one can make measurements in any arbitrary basis $\{\ket{0_{\nu}},\ket{1_{\nu}}\}$. 

As an illustrative example, we consider the four measurements introduced by White \emph{et al.} \cite{White1999} which correspond to the four measurement operators  
\begin{subequations}
\begin{align}
\hat\Pi^{0}={}&\ket{0}\bra{0}\,;\\
\hat\Pi^{1}={}&\ket{1}\bra{1}\,;\\
\hat\Pi^{1}_{\nu}={}&\mathcal{\hat{U}}_{\nu}\ket{1}\bra{1}\mathcal{\hat{U}}\dg_{\nu}\,; ~~~\textrm{for}~\nu=\textrm{i, ii}\,,
\end{align}
\end{subequations}
where the unitary operations are given by
\begin{subequations}
\begin{align} 
\mathcal{\hat{U}}_{\mathrm{i}}={}&\mathcal{\hat U}(0,\pi/2,0)\,;\\
\mathcal{\hat{U}}_{\mathrm{ii}}={}&\mathcal{\hat U}(0,\pi/2,\pi/2)\,,
\end{align}
\end{subequations}
and $\mathcal{\hat U}(h_z,h_{c},\phi)$ is defined in Equation (\ref{u}). The Jacobian for this scenario is given by
\begin{align}\label{jacsqt}
J_{\textsc{a}}={}&\text{det}\left(\frac{\partial (n^{0},n^{1},n^{1}_{\mathrm{i}},n^{1}_{\mathrm{ii}})}{\partial (\rho_{00},\rho_{01},\rho_{11}, \gamma)} \right)={}\rho_{01}\,.
\end{align}
As long as  $\hat\rho$ has non-zero coherence, we can solve for all unknown parameters $\Gamma=\{\rho_{00},\rho_{01},\rho_{11}, \gamma\}$. If $\rho_{01}=0$, the phase  $\gamma$ will be undefined; however, the system of equations will still be invertible as in this situation $\gamma$ will be absent from the expressions for $n^{j}_{\nu}$.

\subsection{One unknown parameter in the unitary}
We now consider a scenario where  $\hat{G}$ contains one unknown parameter. To solve for this additional parameter, we require one additional  measurement. We consider the scenario introduced by Bra\'nczyk \emph{et al.} \cite{aggie11} (with some minor adjustments), which requires ${h}_{z}=0$, and ${h}_{c}={{m}_{c}}{\lambda_{c}}$ where ${{m}_{c}}$ is a controllable parameter and ${\lambda_{c}}$ is unknown. We also require the ability to control $\phi$. Such a physical scenario can be realized when a polarization encoded qubit passes through ${m}_{c}$  wave-plates of unknown retardance ${\lambda_{c}}$ at an orientation specified by $\phi$.  

Under these conditions, we construct five measurement operators:
\begin{subequations}\label{eq:measB}
\begin{align}
\hat\Pi^{1}={}&\ket{1}\bra{1}\,;\\
\hat\Pi^{1}_{\nu}={}&\mathcal{\hat{U}}_{\nu}\ket{1}\bra{1}\mathcal{\hat{U}}\dg_{\nu}\,; ~~~\textrm{for}~\nu=\textrm{i, ii, iii, iv}\,,
\end{align}
\end{subequations}
where the unitary operations are given by
\begin{subequations}
\begin{align}\label{us}
\mathcal{\hat U}_{\mathrm{i}}&={}\mathcal{\hat U}(0,{\lambda_{c}},0)\,;\\
\mathcal{\hat U}_{{\mathrm{ii}}}&={}\mathcal{\hat U}(0,2 {\lambda_{c}},0)\,;\\
\mathcal{\hat U}_{\mathrm{iii}}&={}\mathcal{\hat U}(0,{\lambda_{c}},\pi/2)\,;\\
\mathcal{\hat U}_{{\mathrm{iv}}}&={}\mathcal{\hat U}(0,2 {\lambda_{c}},\pi/2)\,,
\end{align}
\end{subequations}
and $\mathcal{\hat U}(h_z,h_{c},\phi)$ is defined in Equation (\ref{u}). 
 
The Jacobian for this scenario is given by
\begin{align}
J_{\textsc{b}}={}&\text{det}\left(\frac{\partial (n^{1},n^{1}_{\mathrm{i}},n^{1}_{\mathrm{ii}},n^{1}_{\mathrm{iii}},n^{1}_{\mathrm{iv}})}{\partial (\rho_{00},\rho_{01},\rho_{11}, {\lambda_{c}}, \gamma)} \right)\\
\begin{split}\label{eq:JB}
={}&64 \rho _{01}^2 \sin ^6\left(\frac{{\lambda_{c}}}{2}\right) \cos^4\left(\frac{{\lambda_{c}}}{2}\right)\\
&\times (\sin (\gamma )-\cos (\gamma )) \,.
\end{split}
\end{align}
$J_{\textsc{b}}$ will be equal to zero in the following circumstances:
\begin{subequations}
\begin{align}\label{eq:const1}
\rho_{01}&={}0\,;\\\label{eq:const2}
{\lambda_{c}}&={}0\,;\\\label{eq:const3}
{\lambda_{c}}&={}\pi/2\,;\\\label{eq:const4}
\gamma&={}\pi/4\,,
\end{align}
\end{subequations}
and  therefore the transformation will not be invertible. Let's examine each circumstance individually. The case where $\rho_{01}=0$ was discussed in the previous section. When ${\lambda_{c}}=0$, we never change the measurement basis and thus trivially nothing can be learnt about the state except for the value of $\rho_{00}$. If ${\lambda_{c}}=\pi$, then $\mathcal{\hat{U}}_{\mathrm{ii}}=\mathcal{\hat{U}}_{\mathrm{iv}}$, leading to redundancy in the measurement statistics thus providing insufficient information. To understand the last condition, one must look at the explicit forms of ${n}^{1}_{\mathrm{i}}$ and ${n}^{1}_{\mathrm{ii}}$, which differ from ${n}^{1}_{\mathrm{ii}}$ and ${n}^{1}_{\mathrm{iv}}$ by  $\sin(\gamma)\rightarrow \cos(\gamma)$ respectively. When $\gamma=\pi/4$, $\sin(\gamma)= \cos(\gamma)$, once again leading to redundancy in the measurement statistics.  

In all other circumstances, the system will be solvable for the unknown parameters in the density matrix and the unitary.

\subsection{Two unknown parameters in the unitary}

We now introduce an additional unknown parameter in the generator, such that ${h}_{z}={m}_{z}{\lambda_{z}}$ where ${m}_{z}$ is controllable and ${\lambda_{z}}$ is unknown. We can build upon the results of the previous section, where we had implicitly set ${m}_{z}=0$,  and now make an additional measurement given by the measurement operator
 \begin{align}
\hat\Pi^{1}_{\mathrm{v}}={}&\mathcal{\hat{U}}_{\mathrm{v}}\ket{1}\bra{1}\mathcal{\hat{U}}\dg_{\mathrm{v}}\,,
\end{align}
where the unitary operation is given by
\begin{align}\label{us1}
\mathcal{\hat U}_{\mathrm{v}}&={}\mathcal{\hat U}({\lambda_{z}},0,0)\,.
\end{align}
 The Jacobian for this scenario is given by
\begin{align}\label{jac2}
J_{\textsc{c}}={}&\frac{\partial n_{\mathrm{v}}^{1}}{\partial {\lambda_{z}}}J_{\textsc{b}}={} 6 {\lambda_{z}}^2 \rho_{00}J_{\textsc{b}}\,,
\end{align}
where $J_{\textsc{b}}$ is defined in Equation (\ref{eq:JB}). The Jacobian adopts this simple form because $J_{\textsc{c}}$ is upper block triangular where the first block is identical to $J_{\textsc{b}}$ and the other block is simply given by $\partial n_{z}^{1}/\partial {\lambda_{z}}$. As long as $J_{\textsc{b}}\neq 0$, and ${\lambda_{z}}\neq 0$ in the sixth measurement, the system will be solvable for the unknown parameters in the density matrix and the unitary.

\subsection{Uniqueness of the solution}\label{unique}

The solutions to Equations (\ref{bitstats0}) and  (\ref{bitstats1}) are not always unique. As can be seen from Equations (\ref{functionsf}) and (\ref{functionsg}),  the measurements statistics do not depend on $\phi $ and $\gamma$ independently but rather on their difference $\beta=\phi-\gamma$. Given solutions  for $\gamma$ and $\phi$, there are other solutions $\gamma+\eta$, $\phi+\eta$ that will give rise to the same measurement statistics. Another way to see this it to notice that one can perform a non trivial unitary transformation 
\begin{eqnarray}\label{unitV}
\hat{\mathcal{V}}(\eta)=\ket{0}\bra{0}+e^{i \eta}\ket{1}\bra{1}
\end{eqnarray}
to the state $\hat \rho'$ and to the generator of rotations $\hat G$ that leaves the measurement statistics untouched. To this end, notice that $\hat{\mathcal{V}}$ has no effect on the unrotated projectors, i.e. $\mathcal{V}^\dagger\ket{i}\bra{i}\mathcal{V}=\ket{i}\bra{i}$  for $i=0,1$. Because of this identity, it follows that Equations (\ref{bitstats0}) and (\ref{bitstats1}) can be rewritten as:
\begin{align}
n^i_{\nu}={}&\text{tr}\left(\hat \rho' \hat\Pi_{\nu}^{i}\right)\\
={}&\text{tr}\left( \hat \rho'  \hat{\mathcal{U}_{\nu}}^\dagger \ket{i}\bra{i} \hat{\mathcal{U}_{\nu}} \right)\\
={}&\text{tr}\left( \hat \rho'  \hat{\mathcal{U}_{\nu}}^\dagger \hat{\mathcal{V}}^{\dagger} \ket{i}\bra{i} \hat{\mathcal{V}} \hat{\mathcal{U}_{\nu}} \right)\,.
\end{align}

One can insert the identity $\mathbb{I}=\hat{\mathcal{V}} \hat{\mathcal{V}}^\dagger=\hat{\mathcal{V}}^\dagger\hat{\mathcal{V}}$  on both sides of $\rho'$ and use the cyclical property of the trace to write:
\begin{eqnarray}
n^i_{\nu}=\text{tr}\left( \underline{\hat \rho'}  \underline{\hat{\mathcal{U}_{\nu}}^\dagger} \ket{i}\bra{i} \underline{\hat{\mathcal{U}_{\nu}}} \right)\,,
\end{eqnarray}
where:
\begin{subequations}
\begin{align}
 \underline{\hat \rho'}={}&\hat{\mathcal{V}}\hat{\rho}(\rho_{00},\rho_{11},\rho_{01},\gamma) \hat{\mathcal{V}}^\dagger\\
 ={}&\hat{\rho}(\rho_{00},\rho_{11},\rho_{01},\gamma+\eta)\,
 \end{align}
 \end{subequations}
 and
 \begin{subequations}
 \begin{align}
\underline{\hat{\mathcal{U}_{\nu}}}={}&\hat{\mathcal{V}} \exp(-i\hat{G}(h_z,h_c,\phi)) \hat{\mathcal{V}}^\dagger\\
={}&\exp(-i\hat{G}(h_z,h_c,\phi+\eta))\,.
\end{align}
\end{subequations}

Thus, the multiple solutions found when Equations (\ref{bitstats0}) and (\ref{bitstats1}) are inverted are related by a simple unitary of the form (\ref{unitV}). One simple way to make the inversion single-valued is to demand that the generator of rotations $\hat G$ has positive off-diagonal elements, which is equivalent to applying $\hat{\mathcal{V}}(-\phi)$. Finally, notice that if one only demands real off-diagonal elements then one will find two possible solutions sets that are related by the unitary $\hat{\mathcal{V}}(\pi)=\hat\sigma_z$ as was noticed in \cite{aggie11}.

\section{Generalization to multi-dimensional systems}\label{sec:gen}

\begin{figure}
\begin{center}
\includegraphics[width=\columnwidth]{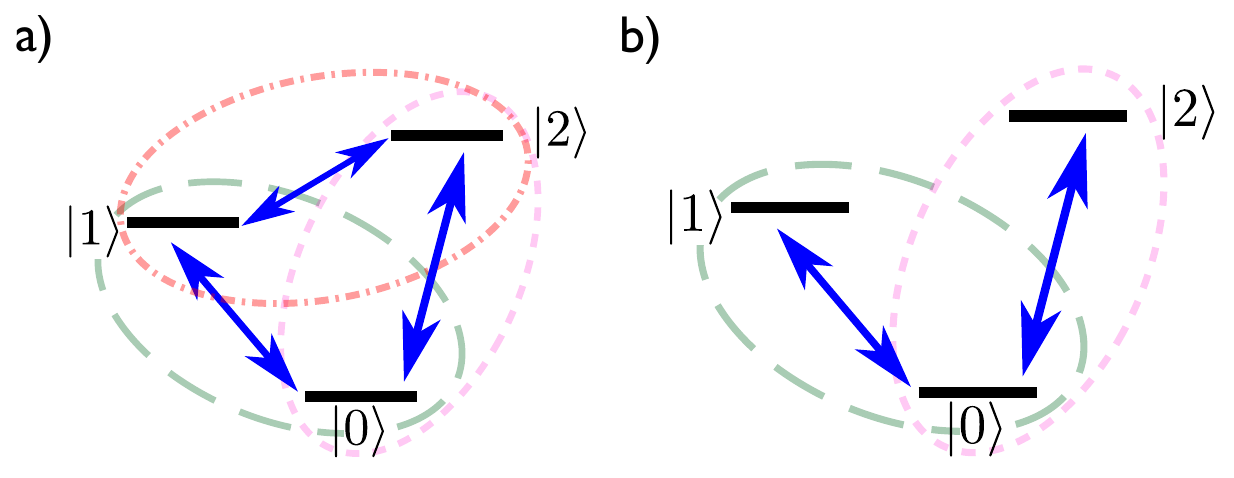} 
\end{center}
\caption{\label{quditsfig}Representation of two qudit systems with $d=3$. In (a) each level can be coupled to each other level by the Hamiltonian interactions represented by the blue arrows. In (b) the two levels $\ket{1}$ and $\ket{2}$ cannot be coupled directly but can still interact to second order through the level $\ket{0}$.}
\end{figure}

Now that we have established the formalism for two-level systems, the generalization to multi-dimensional systems follows readily. One applies the same procedure: 1) write down the unitary being applied to the state, making note of  parameters that are unknown as well as parameters that are controllable; 2) construct a set of measurement operators analogous to those in Equations (\ref{eq:measB}) using different realizations of the unitary, achieved with different settings of the controllable parameters; 3) check that the measurement operators are suitable by calculating the Jacobian in Equation (\ref{eq:jac}); 4) collect measurement statistics from measurements given by the measurement operators in step 2; 5) solve for the unknown parameters using one of a variety of methods, including simple linear inversion, least-squares estimation or the popular maximum likelihood estimation method \cite{Hradil1997} which was used in the original realization of SCT in \cite{aggie11} and gave excellent results in the experiment. Alternatively, one of a growing number of new techniques \cite{Kaznady2009,Blume-Kohout2010,Gross2010a,Toth10,Teo2011,Blume-Kohout2010a,Ng2012,Christandl2011,Blume-Kohout2012,Moroder2012}  can be used.

The above procedure may be simplified if the multi-dimensional system supports interaction Hamiltonians that generate all combinations of pair-wise coupling between the eigenstates, as illustrated for a three-level atomic system in FIG. \ref{quditsfig}(a). In this situation, each measurement operator can be constructed from a two-dimensional unitary operation, and the SCT procedure can be broken down into performing a succession of two-qubit SCT on all possible pairs of levels in the system, denoted by ovals in FIG. \ref{quditsfig}(a). 

On the other hand it might happen that there are eigenstates that are not directly coupled, but are coupled to another common state. This is  the case illustrated in FIG. \ref{quditsfig}(b). In this scenario, pairwise qubit tomography will not be sufficient. Nevertheless it is still possible to gain complete information about the state and, in particular, the value of the coherences between the levels that are not directly coupled, by employing ancillary levels such as the $\ket{0}$ state in FIG. \ref{quditsfig}(b). This will be demonstrated in the next section, for a three-level atomic system in the V configuration. 

We note that the general idea of how to perform quantum state topography of $d$-level systems have been studied by Thew \emph{et al.} \cite{Thew02}.

\section{Example: V-type three-level system}\label{sec:vlev}
In this section we provide an example of how to extend SCT to the case of a three-level system in which two of the levels are not directly coupled by the Hamiltonian that is used to perform changes in the measurement basis; FIG \ref{quditsfig} (b). We consider a concrete physical system, nevertheless the conclusions reached in the following subsections will be independent of the implementations that we outline here. Tomographic reconstruction of three-dimensional systems, given completely characterized unitary operations, has been considered in references \cite{Hofmann2004,Bogdanov2004,Klimov2008}. Here we consider the case where the unitary is not completely characterized.

\subsection{Description of the system}

We consider an atomic or molecular three-level system with a non-degenerate V-type energy ladder with excited states $\ket{1}$ and $\ket{2}$ and ground state $\ket{0}$. The matter Hamiltonian is given by:
\begin{align}
\hat H_{0}=\hbar \omega_1 \ket{1}\bra{1}+\hbar \omega_2 \ket{2}\bra{2}\,,
\end{align}
where the energy of the ground state $\ket{0}$ has been set to zero. The system will couple to the electromagnetic field via the transition dipole moment operator $\hat d$, where 
\begin{subequations}
\begin{align}
\vec{d}_{11}={}&\vec{d}_{22}=\vec{d}_{12}=\vec{d}_{21}=0\,;\\
\vec{d}_{01}={}&\vec{d}_{10}^*\neq 0\,;\\
\vec{d}_{02}={}&\vec{d}_{20}^*\neq 0\,.
\end{align}
\end{subequations}

 The light-matter interaction will be given in the dipole approximation by
\begin{align}
\hat H_{\mathrm{lm}}(t)=\hat d \cdot \vec E (t)\,,
\end{align}
where $\vec E(t)=\vec E^+(t)+\vec E^-(t)$ is the electric field. In particular, it is assumed that the field is in a superposition of two continuous-wave lasers that are each resonant with the $\ket{0}\rightarrow\ket{1}$ and $\ket{0}\rightarrow\ket{2}$  transitions:
\begin{align}\label{Efields}
\vec E^+(t)&={} E_1  e^{i \theta_1}e^{-i\omega_1 t}\vec u_1+E_2  e^{i \theta_2}  e^{-i\omega_2 t} \vec u_2\,,
\end{align}
where $E_i$ and $\theta_i$ are the controllable amplitudes and phases of the fields respectively, $\vec u_i$ are their polarizations, and $\omega_i$ their frequencies. Without loss of generality, we take $E_i$ to be real. In the ``bare'' basis of the V-type system,  the light-matter interaction Hamiltonian is
\begin{align}
\hat H_{\mathrm{lm}}(t)={}& \vec d \cdot \vec E(t)\\
=& \left(\vec d_{10}\ket{1}\bra{0}+\vec d_{20}\ket{2}\bra{0}+\hc\right)\cdot \vec E(t) .
\end{align}
Making the rotating wave approximation and going into the interaction picture with respect to the Hamiltonian $\hat H_0$, the interaction Hamiltonian is given by
\begin{align}\nonumber
\hat H_{\textsc{i}}(t)=&\Big( E_1  e^{i \theta_1}\vec d_{{1}0}\cdot\vec u_1+E_2  e^{i \theta_2}  e^{i\Delta_{21}t} \vec d_{{1}0}\cdot\vec u_2\Big) \ket{1}\bra{0} \\\nonumber
+{}& \Big( E_1  e^{i \theta_1}e^{i\Delta_{12} t}\vec d_{{2}0}\cdot\vec u_1+E_2  e^{i \theta_2}   \vec d_{{2}0}\cdot\vec u_2\Big)\ket{2} \bra{0}\\
+{}&\hc
\end{align}
where $\Delta_{ij}=\omega_{i}-\omega_{j}$. We now calculate the generator of rotations by assuming that the field is turned on at time $-t/2$ and turned off at time $t/2$:
\begin{align}
\hat G={}&\frac{1}{\hbar}\int_{-t/2}^{t/2}d{t}'\hat H_{\textsc{i}}({t}')\\\nonumber
={}& E_1  t e^{i \theta_1}\vec d_{{1}0}\cdot\vec u_1 \ket{1}\bra{0} +E_2  t e^{i \theta_2}   \vec d_{{2}0}\cdot\vec u_2\ket{2} \bra{0} +\hc
\end{align}
where we also assumed that $t\gg \Delta_{12}^{-1}$ such that $\int_{-t/2}^{t/2}e^{i\Delta_{12} t}=2t\mathrm{sinc}(\Delta_{12}t/2)\approx 0$. 
We can write the generator as
\begin{align}\label{G}
\hat G={}&\frac{h_1}{2}\ee{i\phi_1} \ket{1}\bra{0} +\frac{h_2}{2}\ee{i\phi_2}\ket{2} \bra{0} +\hc
\end{align}
where
\begin{subequations}\label{h}
\begin{align}
h_j ={}&2E_j t |\vec d_{{j}0}\cdot\vec u_j| \,;\\
\phi_j={}&\theta_j+\mathrm{arg}(\vec d_{{j}0}\cdot\vec u_j)\,.
\end{align}
\end{subequations}

Equation (\ref{G}) can also be written in the ordered basis $\{\ket{0},\ket{1},\ket{2} \}$ as
\begin{align}\label{tofig}
 G= \frac{1}{2}\left(
\begin{array}{ccc}
 0 & e^{-i \phi_1 }  h_1 &
   e^{-i \phi_2 }  h_2
   \\
 e^{i \phi_1 }   h_1 & 0
   & 0 \\
 e^{i \phi_2 }   h_2 &
   0 & 0
\end{array}
\right)\,.
\end{align}

The matrix above has both zero trace and zero determinant, and therefore the eigenvalues take the very simple form $\text{spec}(\hat G)=\{-\sqrt{h_1^2+h_2^2}/2,0,\sqrt{h_1^2+h_2^2}/2\}$. 

In the more general case where the electric fields in (\ref{Efields}) are not monochromatic but are time-dependent pulses, and assuming that their temporal duration is much longer than $1/\Delta_{12}$, one finds that
\begin{align}
\begin{split}
\hat H_{\textsc{i}}(t)={}&E_1(t)  e^{i \theta_1}\vec d_{{1}0}\cdot\vec u_1 \ket{1}\bra{0}\\
&+ E_2(t)  e^{i \theta_2}   \vec d_{{2}0}\cdot\vec u_2\ket{2} \bra{0}+\hc
\end{split}
\end{align}

The propagator is now strictly given by $\mathcal{T} \exp\left(-\tfrac{i}{\hbar}\int dt H_{\textsc{i}}(t)  \right)$ where $\mathcal{T}$ is the time ordering operator. Given a Hamiltonian $\hat H_{\textsc{i}}(t)$ that commutes with itself at all times, the propagator reduces to the simpler form $\exp\left(-\tfrac{i}{\hbar}\int dt H_{\textsc{i}}(t)  \right)$. We calculate the commutator of $\hat H_{\textsc{i}}(t)$ at two different times ${t}$ and ${t}'$:
\begin{align}
\begin{split}
\big[ &\hat H_{\textsc{i}}(t), \hat H_{\textsc{i}}(t') \big]={}\vec d_{{1}0}\cdot\vec u_1 ( \vec d_{{2}0}\cdot\vec u_2)^* e^{i (\theta_1 -\theta_2 )}\\
&\times \big(E_1(t) E_2(t')-E_1(t')E_2(t)\big)\ket{1}\bra{2}-\hc\,,
\end{split}
\end{align}
where we note that $E_{j}({t})$ is real. The condition that $[ \hat H_{\textsc{i}}(t), \hat H_{\textsc{i}}(t') ]=0$ can be reached if either $E_1(t)=0$, $E_2(t)=0$ or $E_1(t)=E_2(t)$ which are precisely the conditions considered in the remainder of this paper. As long as these conditions are met and the width of the pulses is much longer than $1/\Delta_{12}$, one can safely disregard the time ordering operator $\mathcal{T}$.

\subsection{Measurement statistics}

We now calculate the measurement statistics associated with projection onto the two excited states of the system. A simple way of parametrizing the density matrix of a 3 level system is as follows:
\begin{align}\label{rho}
\hat \rho=\left(
\begin{array}{ccc}
 \rho _{00} &  \rho _{01} e^{-i \gamma _{01}} &  \rho _{02} e^{-i\gamma _{02}} \\
  \rho _{01} e^{i \gamma _{01}} & \rho _{11} &  \rho _{12} e^{-i \gamma _{12}}\\
  \rho _{02} e^{i \gamma _{02}} &  \rho _{12} e^{i \gamma _{12}} & \rho _{22}
\end{array}
\right)\,.
\end{align}

Note that unlike the qubit case, the constraints that the set of numbers $\rho_{ij}$ will have to satisfy so that $\hat \rho$ is a non-negative linear operator are by no means trivial \cite{qutrit,qutrit2}. 

To perform state tomography on the unknown state $\hat\rho$, we evolve it according to $\mathcal{\hat U}_{\nu}=\exp(-i \hat G)$,  where we assume that the interaction time is short compared with the rate of spontaneous emission. After the evolution is complete, the detection of a spontaneously-emitted photon of frequency $\omega_1$ or $\omega_2$ constitutes a projective measurement of $\hat\rho$ given by the projectors 
\begin{align}
\hat\Pi_{\nu}^{1}={}&\ket{1_{\nu}}\bra{1_{\nu}};~~~\hat\Pi_{\nu}^{2}={}\ket{2_{\nu}}\bra{2_{\nu}} \,,
\end{align}
where $\ket{j_{\nu}}=\mathcal{\hat U}_{\nu}\ket{j}$. The statistics associated with measuring $\hat\Pi_{\nu}^{1}$ are given by
\begin{align}\label{n}
n_{\nu}^{1}&={}\text{tr}\left(\hat \rho'\hat\Pi_{\nu}^{1} \right)=\sum_{i\leq j}^{2} f_{ij}\rho_{ij}\,,
\end{align}
where $\hat \rho'=\mathcal{N}\hat \rho $ and the coefficients are given by
\begin{subequations}
\begin{align}
f_{00}={}& \tilde{h}_1^2 {s}^2 \,;\\
f_{11}={} & \left( \tilde{h}_1^2{c}+\tilde{h}_2^2\right)^2\,;\\ 
f_{22}={}&  \tilde{h}_1^2 \tilde{h}_2^2({c}-1)^2\,;\\
f_{01}={}& 2    \sin \left(\beta_{01}\right)\sqrt{f_{00}f_{11}}\,;\\ 
f_{02}={} & 2  \sin \left(\beta_{02}\right)\sqrt{f_{00}f_{22}}\,; \\ 
f_{12}={} & 2 \cos \left(\beta_{12}\right)  \sqrt{f_{11}f_{22}}\,,
\end{align}
\end{subequations} 
where
\begin{subequations}
\begin{align}
\tilde{h}_{j}={}&{h}_{j}/\Omega\,;\\
{c}={}&\cos(\Omega/2)\\
{s}={}&\sin(\Omega/2)\\
\Omega={}&\sqrt{h_1^2+h_2^2} \\
\beta_{0{j}}={}&\phi _{j}-\gamma _{0,{j}}\,;\\
\beta_{12}={}&\gamma _{1,2}+\phi _1-\phi _2\,.
\end{align}
\end{subequations}

We note that there is a non-linear relationship between the measurement statistics $n^1_{\nu}$ and the parameters $h_i$ and $\beta_{jk}$. Similarly, the statistics associated with measuring $\hat\Pi_{\nu}^{2}$  are given by

\begin{align}
n_{\nu}^{2}&={}\text{tr}\left(\hat \rho'\hat\Pi_{\nu}^{2} \right)=\sum_{i\leq j}^{2} g_{ij}\rho_{ij}\,,
\end{align}
where
\begin{subequations}
\begin{align}
g_{00}={}&  \tilde{h}_2^2{s}^2\,; \\
g_{11}={} &\tilde{h}_1^2 \tilde{h}_2^2 ({c}-1)^2 \,;\\ 
g_{22}={}&  \left(\tilde{h}_1^2+ \tilde{h}_2^2{c}\right)^2\,;\\
g_{01}={}& 2 \sin \left(\beta_{01}\right) \sqrt{g_{00}g_{11}}\,;\\ 
g_{02}={} & 2 \sin \left(\beta_{02}\right) \sqrt{g_{00}g_{22}}\,; \\ 
g_{12}={} & 2\cos \left(\beta_{12}\right) \sqrt{g_{11}g_{22}} \,.
\end{align}
\end{subequations}

From the form of $n_{\nu}^{1}$ and $n_{\nu}^{2}$, it can be seen that just one type of measurement, either  $\hat\Pi_{\nu}^{1}$ or $\hat\Pi_{\nu}^{2}$, is sufficient to access all the elements of the density matrix in Equation (\ref{rho}).  We note that, as in the qubit case, the equations are only invertible for $\beta_{ij}$, but not for the individual quantities  $\gamma_{ij}$ and $\phi_{k}$. Only if the phases $\phi_k$ are known, can the complete state be reconstructed. Similarly, only if the phases $\gamma_{ij}$ are known, can the complete unitary be reconstructed. 

Different realizations of  one of the projectors   $\hat\Pi_{\nu}^{j}$ can be used to construct a set of equations invertible for the unknown parameters of the density matrix, thereby realizing quantum state tomography of the three-level system and/or quantum process tomography of the unitary.

Physically, both fields which couple $\ket{1}{}$ and $\ket{2}{}$ to the ground state must be present simultaneously in order to access all elements of the density matrix. This is to be expected since the only way to have a coherent interaction between the two excited states in a V-type system is by going through the ground state---possible only when both fields are present. This can also be seen from the equations above, since $f_{12}$ and $g_{12}$ are dependent on $f_{11}f_{22}$ and $g_{11}g_{22}$ respectively. 

\subsection{Unknown transition dipole moments}

In this section, we demonstrate how self-calibrating tomography can be performed on a V-type three-level system with unknown transition dipole moments $\vec d_{{j}0}$. Without loss of generality, we assume that the transition dipole moments are real and positive. We can therefore parameterize the generator in Equation (\ref{G}) as follows:
\begin{subequations}\label{h}
\begin{align}
h_j ={}&{m}_{j}\lambda_{j} \,;\\
\phi_j={}&\theta_j\,,
\end{align}
\end{subequations} 
where ${m}_{j}=2E_j t $ is a controllable parameter proportional to the square of the intensity of the fields,  $\lambda_{j}=|\vec d_{{j}0}\cdot\vec u_j|$ is unknown and $\theta_j$ is the controllable phase of each field. 

To demonstrate the process of constructing the appropriate measurement operators, we break the problem up into three parts. The first part consists of treating one ground-excited state pair as a two-level system; this gives 5 measurement operators analogous to the qubit case discussed above. The second part consists of doing the same for the other ground-excited state pair; this gives four additional measurement operators since the ground state population parameter is already obtained from the first part. The third part consists of cleverly picking two more operators to solve for the coherences between the two excited states. 

The first set of measurement operators are given by
\begin{subequations}\label{eq:measqtrit}
\begin{align}
\hat\Pi^{1}={}&\ket{1}\bra{1}\,;\\
\hat\Pi^{1}_{\nu}={}&\mathcal{\hat{U}}_{\nu}\ket{1}\bra{1}\mathcal{\hat{U}}\dg_{\nu}\,; ~~~\textrm{for}~\nu=\textrm{i, ii, iii, iv}\,,
\end{align}
\end{subequations}
where the unitary operations are given by
\begin{subequations}
\begin{align}\label{ust1}
\mathcal{\hat{U}}_{{\mathrm{i}}}&={}\mathcal{\hat U}(\lambda_1,0,0,0)\,;\\
\mathcal{\hat{U}}_{{\mathrm{ii}}}&={}\mathcal{\hat U}(\lambda_1,0,0,\pi/2)\,;\\
\mathcal{\hat{U}}_{{\mathrm{iii}}}&={}\mathcal{\hat U}(2 \lambda_1,0,0,0)\,;\\
\mathcal{\hat{U}}_{{\mathrm{iv}}}&={}\mathcal{\hat U}(2 \lambda_1,0,0,\pi/2)\,,
\end{align}
\end{subequations}
where 
\begin{align}
\mathcal{\hat U}({h}_{1},{h}_{2},\phi_{1},\phi_{2})=\ee{-i \hat G({h}_{1},{h}_{2},\phi_{1},\phi_{2})}\,,
\end{align}
and $G({h}_{1},{h}_{2},\phi_{1},\phi_{2})$ is defined in Equation (\ref{G}). The second set of measurements is given by
\begin{align}\label{eq:measqtrit2}
\hat\Pi^{2}_{\nu}={}&\mathcal{\hat{U}}_{\nu}\ket{2}\bra{2}\mathcal{\hat{U}}\dg_{\nu}\,; ~~~\textrm{for}~\nu=\textrm{v, vi, vii, viii}\,,
\end{align}
where
\begin{subequations}
\begin{align}\label{ust2}
\mathcal{\hat{U}}_{{\mathrm{v}}}&={}\mathcal{\hat U}(0,\lambda_2,0,0)\,;\\
\mathcal{\hat{U}}_{{\mathrm{vi}}}&={}\mathcal{\hat U}(0,\lambda_2,0,\pi/2)\,;\\
\mathcal{\hat{U}}_{{\mathrm{vii}}}&={}\mathcal{\hat U}(0,2 \lambda_2,0,0)\,;\\
\mathcal{\hat{U}}_{{\mathrm{viii}}}&={}\mathcal{\hat U}(0,2 \lambda_2,0,\pi/2)\,.
\end{align}
\end{subequations}

The third set will need to provide information about the coherences between the two excited states. For this, we require both fields to be present simultaneously. 

\begin{align}\label{eq:measqtrit3}
\hat\Pi^{1}_{\nu}={}&\mathcal{\hat{U}}_{\nu}\ket{1}\bra{1}\mathcal{\hat{U}}\dg_{\nu}\,; ~~~\textrm{for}~\nu=\textrm{ix, x}\,,
\end{align}
where
\begin{subequations}
\begin{align}\label{ust3}
\mathcal{\hat{U}}_{{\mathrm{ix}}}&={}\mathcal{\hat U}(\lambda_1,\lambda_2,0,0)\,;\\
\mathcal{\hat{U}}_{{\mathrm{x}}}&={}\mathcal{\hat U}(\lambda_1,\lambda_2,0,0+\pi/2)\,.
\end{align}
\end{subequations}

To verify that the measurement statistics $n^{j}_{\nu}=\mathrm{tr}(\hat\Pi^{j}_{\nu}\hat\rho)$, given by the above operators, generate a set of equations invertible for the unknown parameters, we calculate the Jacobian defined in Equation (\ref{eq:jac}). We note that the three sets of inversions are independent of each other. Therefore, as long as one can retrieve: $\{\rho_{00},\rho_{11},\rho_{01},\lambda_1,\gamma_1\}$ from the first set of measurements in Equations (\ref{eq:measqtrit}); $\{\rho_{22},\rho_{02},\lambda_2,\gamma_2\}$ from the second set in Equations (\ref{eq:measqtrit2}); and $\{\rho_{12},\gamma_{12}\}$ from the third set in Equations (\ref{eq:measqtrit3}) the inversion is possible. 

\begin{figure}
\begin{center}
\includegraphics[width=0.8\columnwidth]{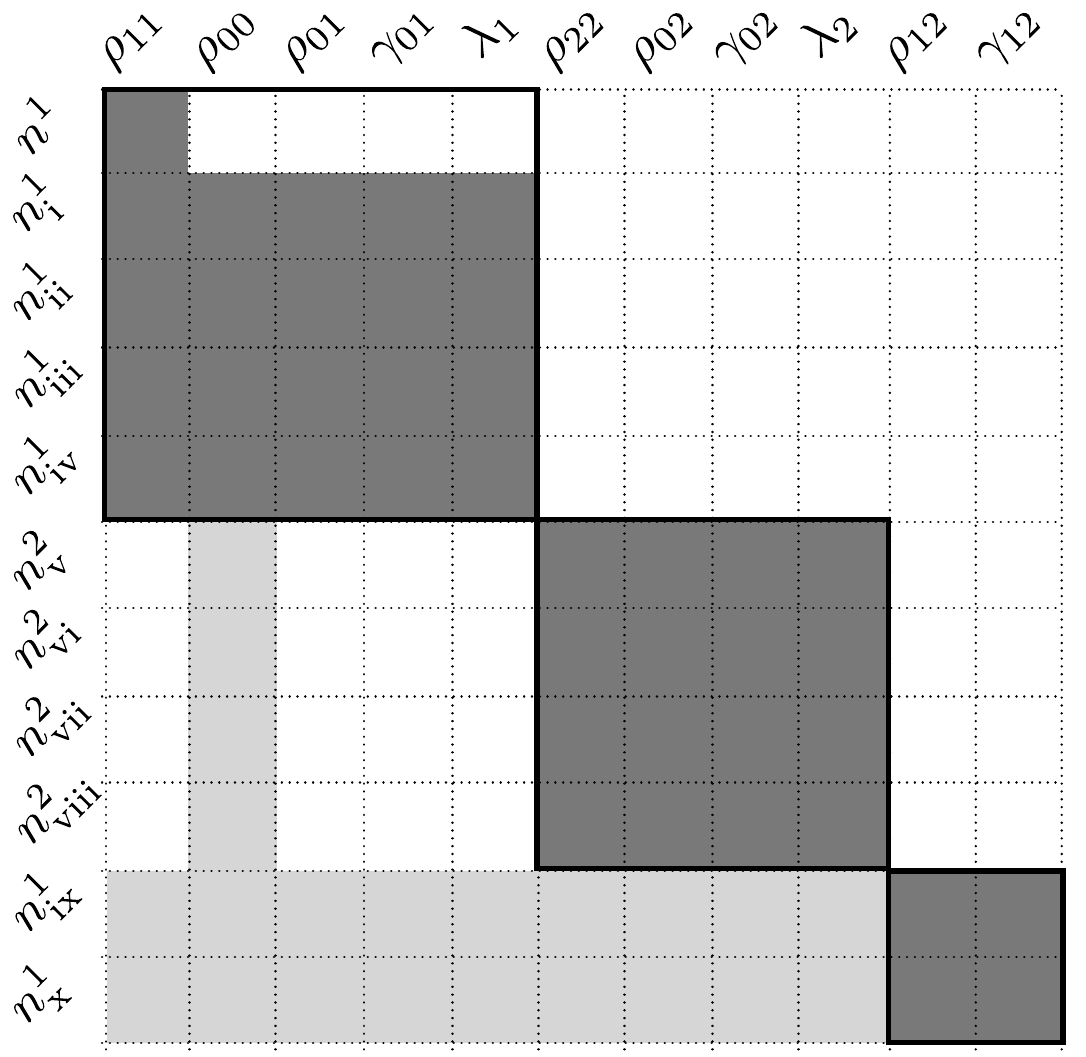}
\caption{\label{blocks}Structure of the Jacobian matrix for the V-type three-level system. White squares indicate a null element of the matrix. Black lines denote the three blocks associated with $J_1, J_2$ and $J_3$. Gray non-zero elements outside the black boundaries signify that each block can only be solved if the blocks before it have already been solved.}
\end{center}
\end{figure}

Equivalently, one can compute the Jacobian matrix and notice that it is upper block triangular where the first $5\times5$ block corresponds to (\ref{eq:measqtrit}), the second $4\times4$  block corresponds to (\ref{eq:measqtrit2}), and the final $2\times2$ block corresponds to (\ref{eq:measqtrit3}). This is demonstrated schematically in FIG. \ref{blocks}. Explicitly the Jacobians for the blocks are:
\begin{subequations}
\begin{align}
\begin{split}
J_1={}&64 \sin ^6\left(\frac{\lambda_1}{2}\right) \cos ^4\left(\frac{\lambda_1}{2}\right) \rho _{01}^2\\
&\times\left(\sin \left(\gamma _{01}\right)-\cos \left(\gamma _{01}\right)\right)\,;
\end{split}\\
J_2={}&2 \sin ^4(\lambda_2) \cos (\lambda_2) \rho _{02}^2 \left(\cos \left(\gamma _{02}\right)-\sin \left(\gamma _{02}\right)\right)\,; \\
J_3={}&-\frac{16 \lambda_1^2 \lambda_2^2 \sin ^4\left(\frac{\Omega }{4}\right) \rho _{12} \left(\lambda_1^2+\lambda_2^2\cos \left(\frac{\Omega }{2}\right)\right)^2}{\Omega ^8}\,.
\end{align}
\end{subequations}

Given the block structure of the matrix, the Jacobian of the whole transformation is simply
\begin{align}\label{eq:jac3}
{J}={J}_{1}{J}_{2}{J}_{3}\,.
\end{align}

The analysis of the singularities of the quantities above (i.e. when $J_i=0$) is completely analogous to that for the qubit case. In fact, $J_1=J_B$ in Equation (\ref{eq:JB}) if $\gamma_{01}\rightarrow \gamma$ and $\lambda_1\rightarrow\lambda_{c}$. We note that in all cases, the Jacobians are proportional to the coherences simply because the phases $\gamma_{i,j}$ are undefined if the coherences $\rho_{i,j}=0$.

If the Jacobian in Equation (\ref{eq:jac3}) is non-zero, we can perform complete quantum state tomography on the three-level system \emph{and} partial quantum process tomography on the unitary, solving for  the unknown  transition dipole moments of the three-level system.

We also note that the measurement operators presented here are by no means unique. Other unitaries, such as those that always contain both fields, may be preferable from an experimental perspective. The advantage of the above choice of operators is two-fold. Firstly, it demonstrates that the task of constructing measurement operators for larger-dimensional systems can largely be broken up into the less daunting task of performing SCT on two-level systems. Secondly, it allows for a more efficient computation of the Jacobian to test the choice of operators.

This analysis can be extended to systems with more complex coupling structures. One can use perturbation theory to show that if two levels are coupled by $d$ ancillary levels, an expansion to order $d+1$ of the propagator will be able to extract the coherence between those levels. For the case of $d=1$ this reduces to the qutrit case  discussed here.

\section{Discussion and concluding remarks}\label{sec:conc}

We have developed a full theoretical formalism for quantum tomography that treats unknown parameters in the state and unitary process on an equal footing. This treatment is applicable to arbitrary-dimensional systems,  where the unitary operations used to change the measurement basis are not completely characterized. 

To date, the duality between unknown parameters in the state and process has been considered in only a handful of publications; characterization of states and processes is still largely treated independently. We hope that the ideas presented in this paper will provide avenues toward a unified treatment that simplifies experiments by eliminating the need for prior calibration, as well as permits characterization of quantum states where state tomography is currently not possible due to lack of well-characterized unitary operations. 

It is important to note that our formalism applies to unitary evolution. This is a valid assumption for some systems, such as free-space linear-optical experiments or atomic systems that are well-isolated from their environment. However, in general, the process will be non-unitary and described by a positive map. This generalization will be discussed in upcoming work. 

\section{Acknowledgements}

This work was funded by the DARPA (QuBE) program and NSERC. The authors thank Hubert de Guise for helpful comments on the manuscript.

\end{document}